# Nuclear Reactor Safeguards and Monitoring with Antineutrino Detectors

Bernstein, A.* , Wang, Y.†, Gratta, G.†, West, T.*

* *Sandia National Laboratories, Livermore CA 94551*

† *Physics Dept, Stanford University, Stanford CA 94305*

**Cubic-meter-sized antineutrino detectors can be used to non-intrusively, robustly and automatically monitor and safeguard a wide variety of nuclear reactor types, including power reactors, research reactors, and plutonium production reactors. Since the antineutrino spectra and relative yields of fissioning isotopes depend on the isotopic composition of the core, changes in composition can be observed without ever directly accessing the core itself. Information from a modest-sized antineutrino detector, coupled with the well-understood principles that govern the core's evolution in time, can be used to determine whether the reactor is being operated in an illegitimate way. A group at Sandia is currently constructing a one cubic meter antineutrino detector at the San Onofre reactor site in California to demonstrate these principles.**

Reactor safeguards regimes, such as the regime implemented by the International Atomic Energy Agency (IAEA) in accordance with the Non-proliferation Treaty (NPT), are intended to detect illicit or suspicious uses of these facilities. Depending on the regime, examples of illicit use could include unauthorized changes in the rate of plutonium production within a reactor, a reduction in the level of irradiation of fuel to facilitate later removal of fissile material, or as in the case of IAEA safeguards, the actual diversion of fissile material from the reactor.

Safeguards monitoring systems are currently in place at about half of the world's power reactors, and at hundreds of research reactors worldwide[1]. In large part, these reactors are now safeguarded by indirect means that do not involve the direct measurement of the fissile isotopic content of the reactor, but instead consist primarily of semi-annual or annual inspections of coded tags and seals placed on fuel assemblies, and measures such as video surveillance of spent fuel cooling ponds. When direct measurements do take place, they are implemented offline, before or after fuel is introduced into the reactor. These may include the counting of fuel bundles or the checking of the enrichment of random samples of fresh or spent fuel rods.

The technique described here differs from these methods in a fundamental way: it provides real-time quantitative information about the reactor core power and its isotopic composition, from well outside the core, while the reactor is online. The advantages of using antineutrinos for reactor safeguards and monitoring include the availability of real-time information on the status of the core, a possible decrease in the time to detection of unauthorized use or diversion of fissile material, less intrusiveness, and simplified operations from the standpoint of both the reactor operator and the safeguards agency.

While the very small antineutrino interaction cross-section represents a basic limitation in the applications discussed here, we will show that, in the specific area of reactor safeguards, a conventional detector with an active volume of about one cubic meter placed in the vicinity of the core (tens of meters) can provide useful, high-statistics data to supplement or provide an alternative to present monitoring techniques. The wealth of activity in the field of neutrino physics in recent years provides us with well-proven, reliable detection techniques.

## Goals for IAEA and other Safeguards Regimes

The IAEA safeguards regime, implemented as part of the Nuclear Nonproliferation Treaty, is the most widespread and important of existing safeguards regimes. IAEA safeguards seek to establish whether a ``significant quantity" of fissile material has been diverted with the ``conversion time". The definition of significant quantity of material depends on the isotope and its physical form, and provides a rough measure of the amount of material needed to manufacture a nuclear weapon. The conversion time is the IAEA estimate of the time it would take to manufacture an actual weapon, taking into account the physical form of the fissile material in question. For plutonium in partially irradiated or spent fuel as found in reactor cores, the conversion time is approximately 1-3 months, and the significant quantity is 8 kg. Antineutrino detection using current technology can approach these goals for reactors.

Though presently not applied, perhaps for lack of enabling technologies, it is easy to envision additional useful goals for reactor safeguards that could enhance IAEA or other regimes. Examples include direct monitoring of the rate of production of fissile material or the degree of irradiation ("burnup") of fuel. The utility of the former measurement lies in ensuring that fissile material is not produced at a rate greater than specified in a safeguards agreement. Tracking changes in burnup rates is also useful, since reducing burnup can facilitate recovery of fissile material (albeit usually at the expense of a reduction in the amount of fissile material available for recovery). Antineutrino detection offers a unique, non-intrusive, real-time method for accomplishing these goals.

## Range of Application

There were 438 operating power reactors worldwide at the end of 2000[2]. Most are Light Water Reactors (LWR) fueled with Low-Enriched Uranium (LEU). About 50 are heavy water reactors (HWR) fueled with natural uranium, primarily of the Canadian Deuterium Uranium (CANDU) design. Both types of reactors have power ratings ranging from several hundred to about 3000~MWt. There are also about 600 research reactors worldwide, with great variations in fueling, power ratings and overall design. Most power reactors and many research reactors have high enough power and thus antineutrino flux to be amenable to safeguarding with antineutrino detectors[3].

The design of particular antineutrino safeguards system will vary according to the reactor power rating, refueling strategy, physical infrastructure and other factors. For example, an independent power measurement may be required to fully exploit the antineutrino signature as a safeguards tool in some cases. Reactors that are refueled online, such as CANDUs, can be continuously monitored for unusual changes in the antineutrino count rate. Additional information about startup parameters might be needed to gauge the significance of such changes for offline-refueled reactors such as LWRs.

**Reactor Safeguards with Antineutrino Detectors**

As a reactor core proceeds through its irradiation cycle, the mass of each isotope varies in time. Initially, only uranium is consumed by fission, while plutonium gradually builds up and begins to fission as well. In the course of a year or more, the total fissile inventory is slowly reduced until the reactor must be refueled. Hence the relative fission rates of the isotopes vary significantly throughout the reactor cycle, even when constant power is maintained.

Antineutrino emission in nuclear reactors arises from the β-decay of neutron-rich fragments produced in heavy element fissions. Over the last two decades, many precision measurements and calculations of the β spectra have been performed[4,5,6,7,8,9] in support of neutrino oscillation experiments which, in turn, have directly measured reactor antineutrino spectra. Current experiments have reached accuracies of a few percent on the absolute antineutrino measurement, (including both reactor flux uncertainties and detector systematics).

In general, the average fission is followed by the production of about 6 antineutrinos that emerge from the core isotropically and without attenuation. A common choice for detection is the (relatively) high probability inverse β -decay reaction:

$$\bar{\nu}_e + p \rightarrow e^+ + n \,. \qquad (1)$$

Here the antineutrino ($\bar{\nu}$) interacts with free protons (p) present in the detection material. The neutron (n) and positron (e+) are detected in close time coincidence, providing a dual signature that is robust with respect to the backgrounds that generally occur at the few MeV energies characteristic of these antineutrinos. In addition to the antineutrino flux, the reaction (1) also allows measurement of the antineutrino energy as:

$$E_{\bar{\nu}} = E_e - M_p + M_n + m_e + O(\frac{m_e}{M_p}) \qquad (2)$$

where $E_{\bar{\nu}}$ is the antineutrino energy, $E_e$ is the positron kinetic energy, $M_p, M_n$ and $m_e$ are the proton, neutron and electron masses respectively, and $O(\frac{m_e}{M_p})$ are terms of order $\frac{m_e}{M_p}$ that mainly account for the nuclear recoil. The reaction in (1) has an energy-dependent cross-section and a threshold of ~1.81 MeV.

Figure 1:The detected $\bar{\nu}$ relative energy spectra. Relative energy spectra of $\bar{\nu}$ produced by the fission of $^{239}$Pu and $^{235}$U, multiplied by the energy-dependent cross-section, $\sigma_{\bar{\nu}}$. The reaction threshold is ~1.81 MeV.

Figure 1 shows the energy spectrum of fission antineutrinos, folded with the energy-dependent detection cross section of (1) for the two most important fissile elements $^{239}Pu$ and $^{235}U$. Because of the spectral differences between these elements, the total $\bar{\nu}$ rate measured in the 2-8 MeV range will change significantly as a function of time, even at constant power. This rate is shown in Figure 2 for an LEU-fueled reactor (2.7% enriched uranium) along with the individual rates for $^{239}Pu$ and $^{235}U$. (Of course, only the total rate is actually measured by a detector). The antineutrino rate is fixed by the initial enrichment and the neutron flux in the reactor. The latter quantity is adjusted with control rods and soluble neutron poisons to regulate criticality and maintain thermal power levels (here assumed constant). Though not shown here, contributions from other isotopes such as $^{238}U$ and $^{241}Pu$ can be accurately accounted for and do not change our results. Similar results are also obtained with diverse fuel and reactor types.

Figure 2: The antineutrino event rate as a function of time. the relative rate of $\bar{\nu}$ events for $^{239}Pu$, $^{235}U$ and the sum of both, as a function of time, from a representative LEU fueled power reactor.

As a rough figure of merit, for the reactor simulated here, a switch of a given weight percentage of plutonium in the core to uranium (relative to total fissile content) induces a change in the antineutrino count rate of about half this percentage. Thus, in the largest commercial power reactors where the fissile content is approximately 3000 kg, a measurement accuracy on the antineutrino count rate at below the percent level is required to detect diversion of an IAEA significant quantity of 8 kg of plutonium within the 1-3 month conversion time for unseparated plutonium. Smaller commercial reactors, research and plutonium production reactors may have fissile inventories five to ten times smaller or less, so that removal of a significant quantity of plutonium would require sensitivity to only a several percent change in fissile content.

**Detector sizes and expected rates**

The detection of MeV-energy antineutrinos via the reaction (1) has been standard in nuclear physics since the early experiments leading to the discovery of the neutrino[10]. In the past 45 years many detectors have been optimized and built in order to reduce backgrounds in very large fiducial masses and to improve the precision of the measurements. Modern detectors, such as Chooz[11] and Palo Verde[12], have fiducial masses of several tons and have run for a few years with total detector-related systematic errors on the antineutrino count rate as low as 1.5%. Without being explicitly designed for the purpose, these detectors have run unattended for weeks. There are few apparent obstacles to unattended operation for months or longer, as would be desirable for safeguards and monitoring purposes. Active media are organic scintillators based on solvents such as trimethylbenzene or isopropylbenzene mixed with paraffin oils, with very good time stability, compatibility with plastic hardware and modest health hazards. For the small-sized detectors needed for nuclear safeguards it is possible to use completely bio-compatible scintillators based on phenylxylylethane or di-isopropylnapthalene, or to use blocks of solid plastic scintillator.

In the process (1) the ionization signal from the $e^+$ and its annihilation is followed by neutron capture, providing, as mentioned, a delayed coincidence that substantially reduces backgrounds. The neutron capture time is 170 μsec in simple organic scintillators, where the dominant process is $n + p \to d + \gamma(2.2\,\mathrm{MeV})$.

Lower backgrounds can be achieved by doping the scintillators with elements having large neutron capture cross-sections, such as Gd, Cl or Li. Scintillators loaded with 0.1% Gd[11,12] have a tight time coincidence (30 μsec) and a higher energy (~8 MeV) capture signature.

Using the well-known cross section for (1) we can estimate the antineutrino event rate in a 1000 kg mass of scintillator at 24 m distance from the core of a 3.4~GWt power reactor. The detector standoff distance will of course differ depending on the reactor site: the example here is the distance of a detector now under construction at the San Onofre nuclear reactor in Southern California. Other reactors may have shorter standoff distances. We obtain an interaction rate of 6850 per day and, using a 40 detection efficiency extracted from Monte Carlo simulations, we are left with a measured count rate of 2740 events per day. Our efficiency estimate is consistent with that reported for real homogeneous detectors[11].

Figure 3: A representative antineutrino detector for safeguards. Schematic view of a possible antineutrino detector for safeguards. Shielding and veto counters are not shown.

As shown in Figure 3, a practical detector could consist of a one meter cube of liquid scintillator contained in an acrylic vessel. To simplify the readout, light from primary scintillation could be absorbed in wavelength shifting plates and re-emitted (at larger wavelength) so that full collection can be achieved with a few photomultipliers (PMT's). In our simulation a respectable light yield of ~150 photoelectrons/MeV is obtained with 4 5-inch PMT's. The antineutrino detection volume can be sufficiently isolated from external gammas and neutrons using a ~50 cm thick layer of water or borated polyethylene. A plastic scintillator envelope, similar in shape to the wavelength shifters can be used to identify and reject cosmic rays crossing the detector.

Very accurate predictions of the efficiency and backgrounds in this type of detector can be obtained using simulation packages such as GEANT[13] together with GFLUKA[14] (hadronic interactions) and GCALOR[15] (n transport). With the few-meter concrete shielding from the reactor reducing artificial neutron backgrounds to ambient levels, we find that the main background arises from cosmic-muon induced spallation neutrons that present the same time coincident signature as the antineutrino events. For a 2~m overhead concrete cosmic-ray shielding this background is estimated at 140 per day, or about 5% of the signal. Random coincidences of γ-rays (and neutrons) from natural radioactivity contribute a much lower rate and are neglected here. Finally, in many plants there are clusters of more than one reactor and hence a ‘background’ will arise from the antineutrinos coming from adjacent reactors. In general for multiple reactor plants the distance between different cores is of order 100 m or more, so that this background is small, and can in any case be directly measured by placing an antineutrino detector near each core.

Present low-energy antineutrino detectors have reached accuracies on the absolute efficiency of better than 2.5%. Much of this uncertainty vanishes for relative measurements, such as needed for many of the safeguards goals of interest here. Statistical error scales with time (t) as $1/\sqrt{t}$ , so that for 10000 events per day, (achievable with a cubic meter detector about 10 meters from the core of a 3 GWt power reactor) 1% statistical accuracy is reached with a day of data and 0.1% in 100 days.

## Examples of Specific Safeguards Applications

We now consider an example of how to detect the removal of fissile material from a core using an antineutrino measurement. We will rely on the changes in antineutrino rate occuring when Pu-bearing fuel assemblies are replaced with fresh fuel. For concreteness we assume a core with 1/3 of its fuel assemblies containing fresh LEU, 1/3 containing fuel irradiated for 18 months, and 1/3 containing fuel irradiated for 36 months, as is typical immediately after refueling in a reactor that has reached its equilibrium operating mode. Thus, in this case, even the initial core has significant Pu content. On the $250^{th}$ day of reactor operation, fuel assemblies containing 8 or 30 kg of plutonium are diverted and replaced with fresh fuel assemblies. We chose a simple non-parametric hypothesis test, designed for sensitivity to systematic shifts in data from an expected mean[16], to calculate the probability that the measured antineutrino rate disagrees with the rate predicted by a precise core simulation (assuming no diversion) after a given period of time. Disagreement is taken as evidence that diversion has occurred. The example here was derived assuming a 1 ton detector at 10 m standoff distance from a 3 GWt power reactor. The results are shown in Table 1.

| Fissile Inventory (kg) | Amount of Pu Diverted (kg) | Probability of Detection within the indicated number of days | | |
|---|---|---|---|---|
| | | 30 days | 60 days | 90 days |
| 2700 | 8 | 0.82 | 0.92 | 0.998 |
| 2700 | 30 | >0.999 | >0.999 | >0.999 |

Table 1: The probability that the indicated change in Pu inventory will be detected by a 1 ton detector, at 10 meters standoff from a 3 GWt reactor, within the indicated period of time. The IAEA goal is the detection of 8 kg of Pu in 1-3 months with a 95% probability.

In this example, only the difference between the antineutrino count rate at startup and on subsequent days is measured. This reduces systematic errors but assumes that the initial fuel loading and fission rate in the reactor are precisely known and provided by independent means. Deriving the fission rate from the absolute antineutrino count rate itself is of course also possible, but introduces the few-percent systematic biases already mentioned. Another limitation of the method derives from the fact that gathering of statistics can occur only when the reactor is critical. In practice, criticality is maintained during refueling only in so-called "on-line refueled" reactors. In the more common off-line refueled reactor, a prolonged shutdown would have the effect of delaying the application of our test for detecting diversion, since, for example, diversion could occur at the beginning of a month-long shutdown, while the antineutrino measurement does not commence until startup.

Antineutrino detection can also be used as a non-intrusive way to meet the other goals mentioned earlier, such as detecting gross changes in reactor power levels or changes in fuel design. A reactor operator seeking to acquire plutonium for weapons could choose to increase the reactor power, thereby increasing the plutonium production rate. Alternatively, a hypothetical diverter of material might seek to facilitate plutonium recovery at the expense of production rate by lowering the reactor power level. Finally, an operator could increase the plutonium production rate by adding fuel with an increased fraction of U-238, adjusting other reactor operating parameters to preserve criticality and the gross power level. All of these diversion strategies are detectable, at some level, through measurement of antineutrinos. Taking as an example the first and simplest case of increased power levels, consider a 10~MWt tank-type heavy water research reactor fueled with natural uranium fuel, similar in design to the Japanese JRR-3 reactor[17]. This reactor produces about 2 kg of fissile plutonium per year. A 44 MWt reactor with the same fuel and similar design, such as the German FR-2 reactor, produces about 5 times as much plutonium per year: we can imagine that a reactor of the first type would be covertly run at the power levels of the second and the excess Plutonium used for weapon production. For such research reactors with transverse dimensions of a meter or so, the detector could be located as close as 10 m from the core, so that (in this example) the increased power would change the absolute antineutrino rate from 50 to about 220 per day, providing clear evidence for a change in operations without intrusive inspection of the reactor.

## Conclusions

Our analysis shows that antineutrino detection holds considerable promise as a novel tool for monitoring many types of nuclear reactors. The potential applicability of the method extends to hundreds of reactors worldwide. The advances in design and operation of low-energy antineutrino detectors, together with the experience accumulated in predicting reactor antineutrino energy spectra make the technique simple, cheap and reliable. These important features will be demonstrated by a pilot detector which Sandia National Laboratories' California branch is now installing at the San Onofre reactor in California.

We thank David Spears and Mike O'Connell of DOE-NN-20 for their encouragement and financial support of this work. We also thank Drs. F. Boehm, L. Miller and A. Piepke for stimulating discussion. We are grateful to the Loyd Wright, Tom Kennedy, John Joyce, Ralph Sampson, Nancy Alms and the rest of the staff of the San Onofre Nuclear Generating Station for their generous assistance, which has made it possible for us to deploy our detector. The Sandia authors express their appreciation for the support provided by Charlie Harmon and Mark Grohman at Sandia Albuquerque, and Vipin Gupta, Carolyn Pura, Will Bolton, Larry Brandt and Pat Falcone at Sandia California.

**<corr> Correspondence and requests for materials should be addressed to Adam Bernstein (e-mail: abernst@sandia.gov).**

---

[1] http://www.iaea.org/worldatom/Periodicals/Factsheets/English/safeguards-e.pdf

[2] http://www.iaea.or.at/programmes/a2/oprconst.html

[3] The power ratings and numbers of reactors of each type are taken from the database at http://www.insc.anl.gov

[4] K. Schreckenbach *et al.*, Phys. Lett. B 160 (1985) 325.

[5] A.A. Hahn *et al.*, Phys. Lett. B 218 (1989) 365.

[6] B.R. Davis *et al.*, Phys. Rev. C 19 (1979) 2259.

---

[7] P.Vogel *et al.*, Phys. Rev. C 24 (1981) 1543.

[8] O.Tengblad *et al.* Nucl. Phys. A503 (1989) 136.

[9] H-V. Klapdor-Kleingrothaus and J. Metzinger, Phys. Rev. Lett. 48 (1982) 127.

10 F. Reines and C. Cowan Jr., Phys. Rev. 92 (1953) 830.

11 M. Apollonio et al., Phys. Lett. B 466 (1999) 415.

12 F. Boehm et al., Phys. Rev. D62 (2000) 072002

13 R. Brun et al., 'GEANT 3', CERN DD/EE/84-1 (revised), 1987.

14 P.A. Aarnio et al., ``FLUKA user's guide'', TIS-RP-190, CERN, 1990.

15 T.A. Gabriel et al., ORNL/TM-5619-mc, April 1977.

16 G.K. Bhattacharyya and R.A. Johnson, ``Statistical Concepts and Methods'' (John Wiley and Sons: New York, 1977), page 519.

17 Nuclear Proliferation and Safeguards (Washington DC, Office of TechnologyAssessment, 1977) Vol. 2, Appendix V, Table B-2, p. V-229.

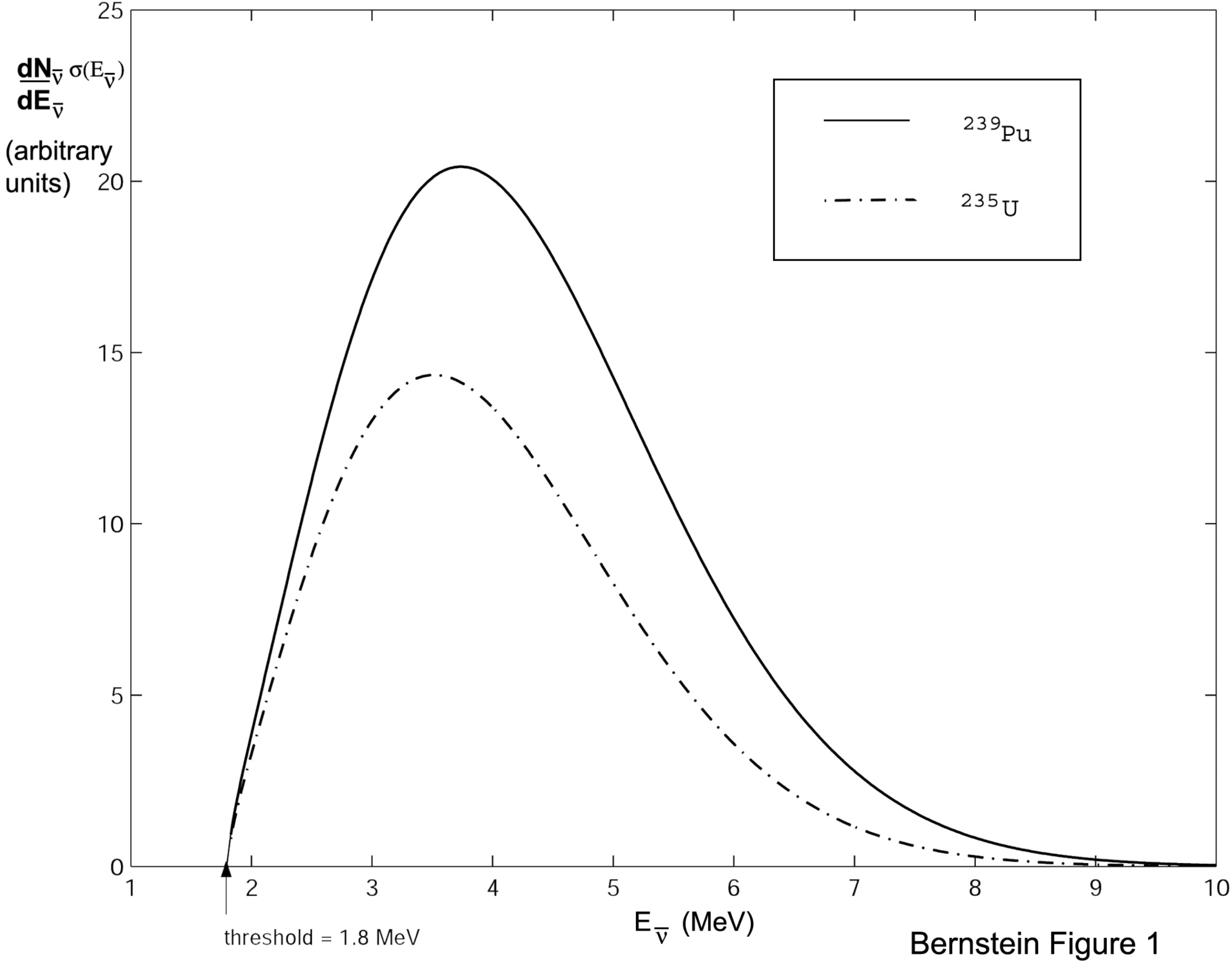

Bernstein Figure 1

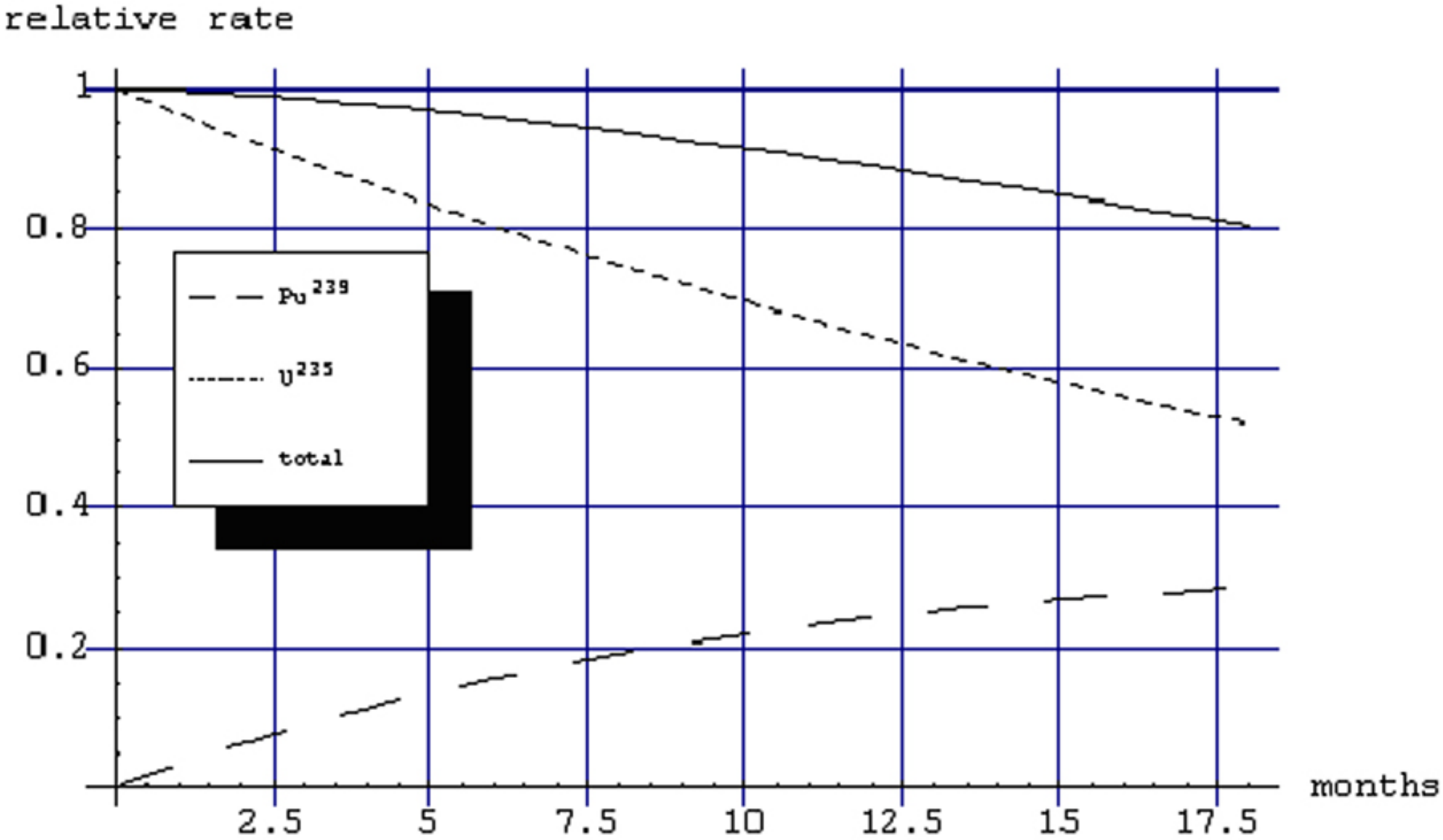

Bemstein Figure 2

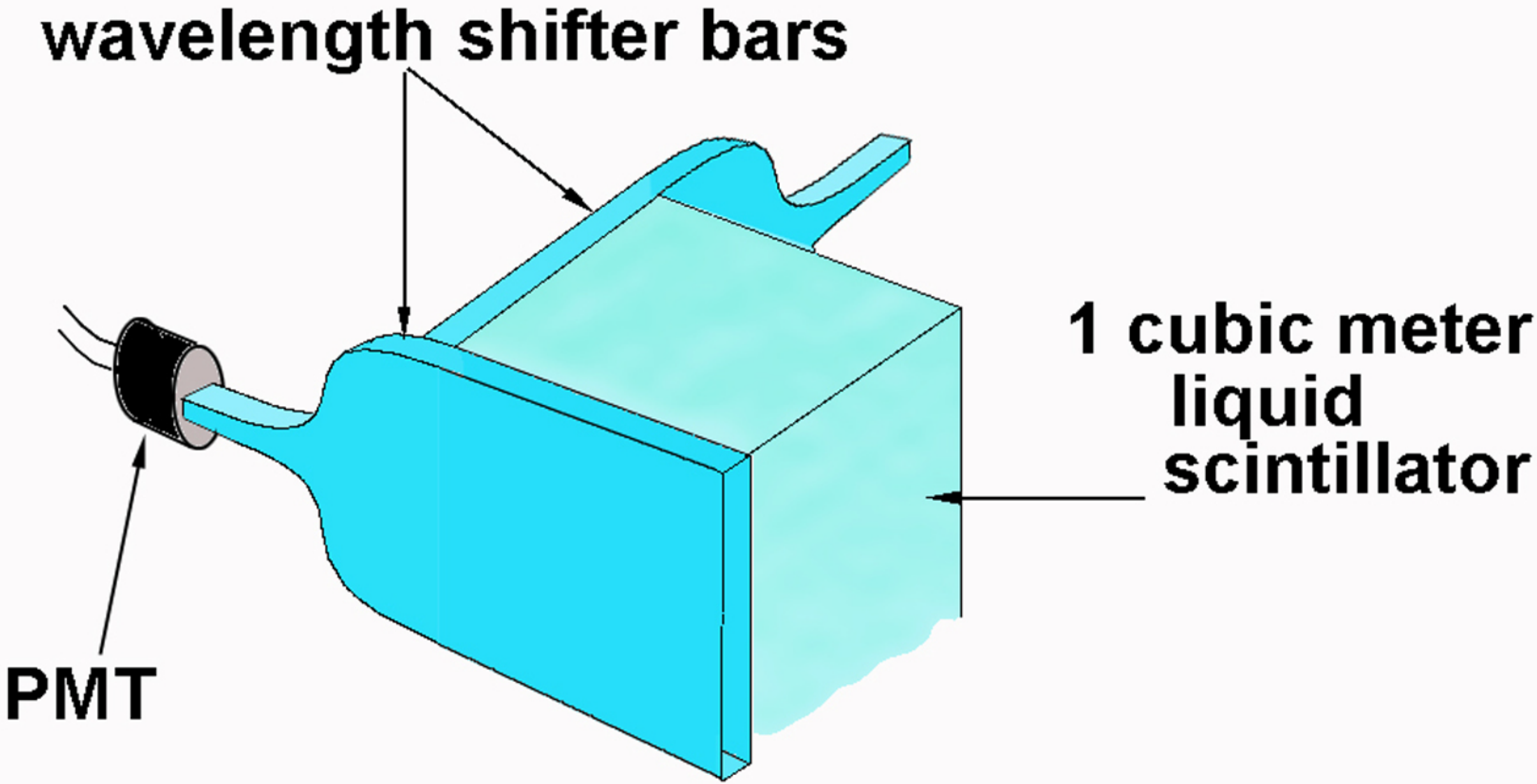

Bernstein Figure 3